\documentclass[
showpacs,
amsmath,amssymb,
aps,
twocolumn,
prb,
floatfix,
nofootinbib
]{revtex4-2}

\usepackage{graphicx}
\usepackage{amssymb}
\usepackage{dcolumn}

\begin{document}

\title{Resonance phenomena in a nanomagnet coupled to a Josephson junction under external periodic drive}

\author{K. V. Kulikov$^{1,2}$,  D. V. Anghel$^{1,3}$, M. Nashaat$^{1,4}$, M. Dolineanu$^{3,5}$, M. Sameh$^{4}$, and Yu. M. Shukrinov$^{1,2,6}$ }

\affiliation{$^1$ BLTP, JINR, Dubna, Moscow region, 141980, Russia\\
	$^2$ Dubna State University, Dubna, Russia\\
	$^3$ \mbox{Horia Hulubei National Institute for R\& D in Physics and Nuclear Engineering, M\u{a}gurele, Romania}\\
	$^4$ \mbox{Department of Physics, Faculty of Science, Cairo University, 12613, Giza, Egypt}\\
	$^5$ \mbox{University of Bucharest, Faculty of Physics, 077125 M\u{a}gurele, Romania}\\
	$^6$  \mbox{Moscow Institute of Physics and Technology, Dolgoprudny 141700, Russia}}

\begin{abstract}
We investigate resonance phenomena in a system consisting of a nanomagnet coupled to a Josephson junction under external periodic drive. The coupling in the system leads to appearance of additional resonance peaks whose properties depend on the periodic signal and Josephson junction dynamics. In the linear regime, we derive an analytical description of the resonance phenomena which is then confirmed by numerical simulations. This analytical method is universal and can be also applied to Josephson junctions with anomalous phase shift in current phase relation. This work provides a new method of controlling the resonances of hybrid structures, which may be interesting for applications.
\end{abstract}


\maketitle

\section{Introduction} \label{sec_intro}

One of the most interesting recent developments in magnetism is the ability to fabricate nanometer-scale magnets \cite{s-s_98, kcw-jap_97, db-s_98, c-s_98, wfa-prb_98, mkghw-apl_98, h-s_96}.
These nanomagnets possess magnetic properties which are different from those of bulk materials and may provide advanced replacements for hard disk media \cite{wnp-ieee_97, w-jvstb_98} and computer memory chips \cite{p-s_98}.
Recently, molecular nanomagnets have also been studied as potential candidates for qubit realization \cite{armbttw-prl_07}. Such a realization is expected to play a crucial role in quantum information processing \cite{ll-n_01} and spintronics \cite{rgblfs-nm_05, bw-nm_08}.

The dynamics of the magnetization of molecular nanomagnets can be described by the Landau-Lifshitz-Gilbert (LLG) equation \cite{hikino2011ferromagnetic,Hillebrands2003}. It is well known that if a material is magnetized by an external magnetic field, the magnetization vector $\textbf{M}$ becomes parallel to the external magnetic field.
When the external magnetic field frequency coincides with the eigenmode of the precession of the magnetic moments of the electronic system of the ferromagnet, a ferromagnetic resonance (FMR) is achieved \cite{Landau1935,hikino2011ferromagnetic,Sharma2013}.
In this case, spin waves are excited in the ferromagnet and can be viewed as both spatial and time dependent variations in the magnetization \cite{hikino2011ferromagnetic,Hillebrands2003}.
Experimentally, FMR was observed by Griffiths~\cite{Griffiths1946}, who found that FMR does not occur exactly at the resonance frequency $\Omega_{r}=\gamma H$, (where $\gamma$ is the electron gyromagnetic ratio and $H$ is the internal magnetizing field).
Furthermore, Kittel proposed that the ferromagnetic resonance condition should be modified from the original Landau-Lifshitz theory by taking into account the shape and the crystalline anisotropy through the demagnetizing fields~\cite{Kittel1947}.
The FMR technique can provide information on the magnetization, magnetic anisotropy, dynamic exchange/dipolar energies and relaxation times, as well as the damping in the magnetization dynamics \cite{Sharma2013}.
In Ref.~\cite{Owens2009} the dynamic fluctuations of the nanoparticles and their anisotropic behavior have been recorded with FMR signal.
The FMR modes for $Fe_{70}Co_{30}$ magnetic nanodots in a mono-domain state under different in-plane and out-of-plane magnetic fields were studied in Ref.~\cite{Miyake2012}.
The FMR technique can be applied to several systems, like monolayers \cite{Zakeri2006}, multilayers, ultrathin films \cite{Layadi2004,Liua2012,Schafer2012}, and nano-systems \cite{Parvatheeswara2006,Wang2011,Patel2012,Biasi2013}.
The direct coupling between the magnetic moment and the Josephson oscillations, realized in Josephson junction (JJ) coupled to a ferromagnet manifests unique properties at the ferromagnetic resonance, such as the appearance of Shapiro-like steps in the IV-characteristics, different stable magnetic trajectories, Duffing oscillator features etc., \cite{hikino2011ferromagnetic, Maekawa2011, Nashaat2018, Nashaat2020, Shukrinov2021}.

Recently, a dramatic increase of the FMR frequency in the presence of electronic interaction between superconducting and ferromagnetic layers, which was due to the coupling of magnetization dynamics and superconducting imaginary conductance at S-F interfaces, was predicted theoretically in Ref.~\cite{Silaev2022}. This have been confirmed experimentally in Ref.~\cite{Golovchanskiy2023}, where the authors consider superconductor-ferromagnet-superconductor thin film structures and observed broad-band ferromagnetic resonance for a large set of samples with varied thickness of both superconducting and ferromagnetic layers in wide frequency and temperature ranges.

A system formed of a nanomagnet coupled to a JJ was analyzed in~\cite{cai2010interaction}. The magnetic field of the nanomagnet influences the superconducting current in the JJ, and vice-versa, the electromagnetic field created by the JJ acts upon the nanomagnet.
Several features are predicted to appear as  a result of this mutual interaction, like, for example, a spin flip, produced by a specific time variation of the external voltage.

The superconducting current of a JJ coupled to an external nanomagnet driven by a time-dependent magnetic field both without and in the presence of an external ac drive are studied in Ref.~\cite{Ghosh2017}.
The authors showed the existence of Shapiro-type steps in the IV- characteristics of the JJ subjected to a voltage bias for a constant or periodically varying magnetic field and explored the effect of rotation of the magnetic field and the presence of an external ac drive on these steps ~\cite{Ghosh2017}. Furthermore, a uniform precession mode (spin wave) could be excited by a microwave magnetic field, at ferromagnetic resonance (FMR), when all the elementary spins precess perfectly in phase~\cite{Lifshitz2003}.

An analogy between the Kapitsa pendulum and the JJ-nanomagnet system was introduced in Ref.~\cite{snrk-jetpl_19} and it was shown that the magnetic moment of the nanomagnet may be reoriented. In this case, the Josephson to magnetic energy ratio $G$ corresponds to the amplitude of the variable force of the Kapitsa pendulum, the Josephson frequency $\Omega_J$ corresponds to the oscillation frequency of the suspension point, and the averaged magnetic moment components specify the stable position.
An analytical description of this analogy and of the observed phenomena was reported in Ref.~\cite{PhysRevB.105.094421.2022.Kulikov}.

In this paper we investigate resonance phenomena in a system consisting of a nanomagnet coupled to a Josephson junction under external periodic drive.
A mechanical analogy for this system are two coupled oscillators with external periodic force applied to one of them.
We show that such a system has a rich resonance physics.

The structure of the paper is as follows. In Section~\ref{sec_model} we introduce the model. The analytical description of the resonance properties of the system is provided in Section~\ref{sec_analytic}.
The comparison of analytical and numerical results, as well as a discussion of some special cases are presented in Section~\ref{sec_discussion}.
The conclusions are presented in Section~\ref{sec_concl}.

\section{Model and Methods} \label{sec_model}

\begin{figure}[b]
	\includegraphics[width=0.52\linewidth]{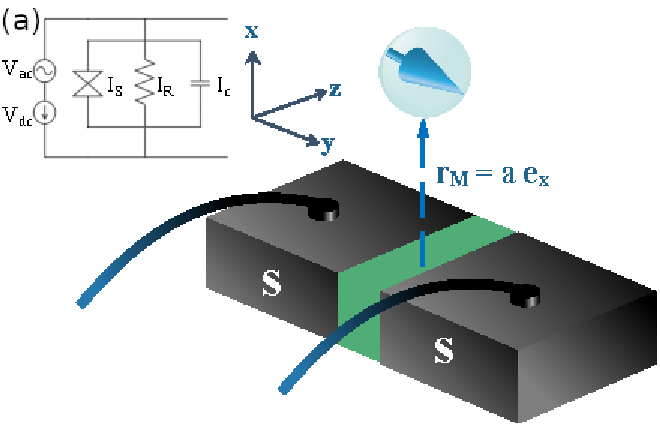}
	\caption{Schematic diagram of the considered system with the system geometry.}
	\label{1}
\end{figure}

A voltage biased JJ, of length $l$, coupled to a nanomagnet of magnetic moment $\textbf{M} = (M_x,M_y,M_z)$ located at distance $\textbf{r}_{M}=a \textbf{e}_x$ from the center of the junction, as shown in Fig.~\ref{1}~(a).
The anisotropy energy of the nanomagnet is given by
\begin{eqnarray}
E_M=\frac{- K v}{2} \bigg(\frac{M_{y}}{M_{s}}\bigg)^{2} ,
\label{anisotropic}
\end{eqnarray}
where  $K$ is the anisotropy constant, $v$ is the volume of nanomagnet, $M_{y}$ magnetization in y-direction (easy axis), $M_{s}$ is the saturation magnetization.
The DC and AC voltages applied on the JJ generate the magnetic field acting on the nanomagnet.
Therefore, the effective field is given by \cite{snrk-jetpl_19, cai2010interaction, Ghosh2017}
\begin{eqnarray}
\textbf{H}_{eff}=-\dfrac{dE}{d\textbf{M}}=-\dfrac{dE_M}{d\textbf{M}}+\dfrac{d}{d\textbf{M}}I \int d\textbf{r} \textbf{A}(\textbf{r},\tau),
\label{Effective_Field}
\end{eqnarray}
where $\textbf{A}=\dfrac{\mu_0}{4\pi}\dfrac{\textbf{M}\times\textbf{r}}{r^3}$ is the vector potential created at a point $r$ from the nanomagnet and assuming that the latter is less than all other dimensions of the problem.
The last term is the magnetic field $H_J$ created by the total current $I$ through the Josephson junction \cite{snrk-jetpl_19}.

The dynamics of the nanomagnet's magnetic moment can be described by Landau-Lifshiz-Gilbert (LLG) equation
\begin{eqnarray}
\dfrac{d\textbf{M}}{d\tau} = \gamma \textbf{H}_{eff} \times \textbf{M}+\dfrac{\alpha}{M_{0}}\left(\textbf{M}\times\dfrac{d\textbf{M}}{d\tau}\right) ,
\label{LLG}
\end{eqnarray}
where $\alpha$ is the Gilbert damping parameter, $M_{0}=\mid\textbf{M}\mid$, $\gamma$ is the gyromagnetic ratio.
In dimensionless quantities the magnetic moment components in LLG equation is given by
\begin{subequations} \label{LLG_components}
\begin{eqnarray}
	\dfrac{dm_{x}}{dt} &=& \frac{\omega_{F}}{(1+\alpha^{2})} \bigg[ h_y \left(m_z-\alpha  m_x m_y\right)
	\label{LLG_components1} \\
	&& - h_z \left(\alpha  m_x m_z+m_y\right)+\alpha  h_x \left(m_y^2+m_z^2\right) \bigg] , \nonumber \\
	\dfrac{dm_{y}}{dt} &=&\frac{\omega_{F}}{(1+\alpha^{2})} \bigg[ -h_x \left(\alpha  m_x m_y+m_z\right)
	\label{LLG_components2} \\
	&& + h_z \left(m_x-\alpha  m_y m_z\right)+\alpha  h_y \left(m_x^2+m_z^2\right) \bigg] , \nonumber \\
	\dfrac{dm_{z}}{dt} &=&\frac{\omega_{F}}{(1+\alpha^{2}) D} \bigg[  \alpha  \tilde{h}_z \left(m_x^2+m_y^2\right) \label{LLG_components3} \\
	&& - h_y \left(m_x+\alpha  m_y m_z\right)+h_x \left(m_y-\alpha  m_x m_z\right) \bigg] ,
	\nonumber
\end{eqnarray}
\end{subequations}
where $m_{i}=M_{i}/M_{s}$ ($i=x,y,z$), are the normalized components of the magnetic moment, $h_{i}$ are the effective field components normalized to $H_{F}=\omega_{F}/\gamma$, $\omega_{F}=\Omega_F/\omega_c$ is the normalized frequency of the ferromagnetic resonance, $\omega_{c}$=$2eRI_{c}/\hbar$ is characteristic Josephson frequency, $I_{c}$ is the critical current of the JJ, $R$ is the resistance of the JJ, $t=\tau\omega_c$ is normalized time, $D=1+\dfrac{\omega_{F} \alpha \epsilon k } {1+\alpha^{2} }\left(  m^{2}_{x} + m^{2}_{y}\right)$, $\epsilon=Gk$,  $G=\epsilon_J / K v$ is the Josephson to magnetic energy ratio, $\epsilon_{J}=\Phi_{0}I_{c}/2\pi$ is the Josephson energy, $\Phi_{0}$ is the flux quantum and $k=(2\pi/\Phi_{0})\mu_{0} M_{s}l /a\sqrt{l^{2}+a^{2}}$ is the coupling constant between the JJ and the nanomagnet.%

The components of the total effective field can be obtained by using the Biot-Savart law to calculate the magnetic field acting on the nanomagnet generated by the Josephson junction. So, they are given by
\begin{eqnarray}
h_{x}&=&0, h_{y} = m_{y},  \nonumber\\ h_{z}& =& \epsilon [\sin(V t - k m_{z}+\frac{A}{\omega_R}\sin(\omega_R t))\nonumber\\&+&  \delta(V+A \cos(\omega_R t) -  k \dot m_{z})]
\label{Effective_Field_components}
\end{eqnarray}
where $V$ is the dc voltage bias normalized to $V_c=\hbar \omega_c/ 2e$, $A=V_{ac}/V_c$ is the amplitude of external drive, $\omega_R$ is the frequency of the external drive normalized to $\omega_c$.
Notice also that in our normalization the dc voltage bias $V$ is equal to the Josephson frequency $\omega_J$.
The parameter $\delta$ takes the values $ 1$ or $0$ and indicates the terms that come from quasiparticle current.
The case with $\delta=0$ is studied in Ref.~\cite{smrbb-epl_18}, where the authors demonstrate the reorientation of the easy axis in $\varphi_{0}$-junction with inverse Kapitsa-like pendulum feature.
Here we consider $\delta = 1$ and take into account the effect of the superconducting and quasiparticle tunneling currents.

\section{Analytical description} \label{sec_analytic}

We consider that both, DC and AC voltages are applied to the JJ, so the nanomagnet is subjected to two periodic drives.
The first one is due to the oscillating magnetic field generated by the Josephson oscillations of the DC voltage biased JJ.
This magnetic field excites the precession of the magnetic moment of the nanomagnet and leads to ferromagnetic resonance when the precession frequency equals the eigenfrequency $\omega_{F}$ of the magnetic system.
The second drive, due to the AC voltage applied to the JJ, leads to the "Kittel" feromagnetic resonance when the AC frequency equals to the eigenfrequency $\omega_{F}$ of the magnetic system.
For experimental realization of such a system, we estimate the model parameters according to Ref.~\cite{Nashaat2022}, based on Refs.~\cite{Mangin2017, Cowburn1999, Yin2006, Buschow2005}.
In this section we present the analytical description of the combined effect of these two drives, which predicts rich resonance physics.

It is easier to work in spherical coordinates, defining the projections of the magnetic moment in terms of the polar and azimuthal angles, namely $m_x=\sin\theta\cos\phi, m_y=\sin\theta\sin\phi, m_z=\cos\theta$, where $\theta\in[0,\pi]$ and $\phi=[0,2\pi]$.
Then, Eqs.~(\ref{LLG_components}) and (\ref{Effective_Field_components}) are translated into time derivatives of the angles, according to the Appendix~A of Ref.~\cite{PhysRevB.105.094421.2022.Kulikov}:
\begin{widetext}
\begin{subequations} \label{derivs_theta_phi_pol}
	\begin{eqnarray}
		\dot\theta &=& - \frac{ \sin\theta \, \Omega_F }{ 1 + \alpha^2 + \delta \alpha \epsilon k \sin^2\theta \, \Omega_F }  \Big[ \alpha \tilde h_z - \sin\phi ( \cos\phi + \alpha \cos\theta \sin\phi) \Big] ,
		\label{deriv_theta} \\
		\dot \phi &=& \frac{ \Omega_F }{ 1 + \alpha^2 + \delta \alpha \epsilon k \sin^2\theta \, \Omega_F } \left[ \tilde h_z
		+ \Big( \sin^2 \theta \cos\phi \delta \epsilon k \Omega_F - \sin\phi \cos\theta + \alpha \cos\phi \Big) \sin\phi \right] , \label{eq_dot_phi_simple} \\
		\tilde h_z(t) &=& \epsilon \sin\left[ V t - k \cos\theta + \frac{A}{\omega_R} \sin(\omega_R t ) \right] + \delta \epsilon \left[V + A \cos(\omega_R t ) \right].
		\label{def_thz}
	\end{eqnarray}
\end{subequations}

In the expression of $\tilde{h}_z(t)$ we write
\begin{subequations} \label{exp_sin_cos_Bessel2}
\begin{eqnarray}
	\sin\left[ V t - k m_z + \frac{A}{\omega_R} \sin\left( \omega_R t \right) \right] &=& \sum_{m=-\infty}^{\infty} \text{sign}^m(m) J_{|m|}\left(\frac{A}{\omega_R}\right)
	\sin \Big[ (V + m \omega_R) t - k m_z \Big] \label{def_g1pg2} \\
	\cos\left[ V t - k m_z + \frac{A}{\omega_R} \sin\left( \omega_R t  \right) \right] &=& \sum_{m=-\infty}^{\infty} \text{sign}^m(m) J_{|m|}\left(\frac{A}{\omega_R}\right)
	\cos \bigg[ ( V + m \omega_R) t - k m_z \bigg] ,
	\label{def_h1ph2}
\end{eqnarray}
\end{subequations}
Plugging Eqs.~(\ref{exp_sin_cos_Bessel2}) into Eqs.~(\ref{derivs_theta_phi_pol}) we obtain an expansion for $\dot{\theta}$ and $\dot{\phi}$:
\begin{subequations} \label{dot_theta_phi}
\begin{eqnarray}
	\dot\theta &=& - C(\theta) \sin\theta   \left[ F_0(\theta, \phi) + F_1(\theta,t) \right] ,
	\label{dot_theta} \\
	\dot \phi &=& C(\theta) \left[ Q_0(\theta,\phi) + Q_1(\theta,t) \right] , \label{dot_phi}
\end{eqnarray}
where
\begin{eqnarray}
	C(\theta) &=& \frac{ \Omega_F }{ 1 + \alpha^2 + \delta \alpha \epsilon k \sin^2\theta \, \Omega_F }\\
	F_0(\theta, \phi) &=& \delta\alpha\epsilon V - \sin\phi ( \cos\phi + \alpha \cos\theta \sin\phi)
	\label{def_F0} \\
	F_1(\theta, t) &=& \delta\alpha\epsilon A \cos(\omega_R t )
	+ \alpha\epsilon \sum_{m=-\infty}^{\infty} \text{sign}^m(m) J_{|m|}\left(\frac{A}{\omega_R}\right) \sin \Big[ (V + m \omega_R) t - k \cos\theta \Big]
	\label{def_F1} \\
	Q_0(\theta, \phi) &=& \delta\epsilon V
	+ \Big( \sin^2 \theta \cos\phi \delta \epsilon k \Omega_F - \sin\phi \cos\theta + \alpha \cos\phi \Big) \sin\phi
	\label{def_Q0} \\
	Q_1(\theta, t) &=& \delta\epsilon A \cos(\omega_R t )
	+ \epsilon \sum_{m=-\infty}^{\infty} \text{sign}^m(m) J_{|m|}\left(\frac{A}{\omega_R}\right) \sin \Big[ (V + m \omega_R) t  - k \cos\theta \Big]
	\label{def_Q1}
\end{eqnarray}
\end{subequations}
\end{widetext}
Notice that the terms $F_0(\theta, \phi)$ and $Q_0(\theta, \phi)$ do not depend explicitly on time.
Therefore, if $F_1(\theta, t) = Q_1(\theta, t) = 0$, and the magnetization of the nanomagnet reaches a position $(\theta_0, \phi_0)$ where
\begin{equation} \label{eqs_stat_point}
	F_0(\theta_0, \phi_0) = Q_0(\theta_0, \phi_0) = 0 ,
\end{equation}
it remains in that position.
We call $(\theta_{0}, \phi_0)$ the \textit{stationary point} or \textit{stationary position}.
From~(\ref{eqs_stat_point}) we get
\begin{equation}
	\phi_0 = \frac{\pi}{2}, \frac{3\pi}{2}, \quad {\rm and} \quad
	\cos\theta_0 =
	\begin{cases}
		\delta \epsilon V & {\rm for} \quad |\delta \epsilon V| \le 1 , \\
		{\rm sign}(V) & {\rm for} \quad |\delta \epsilon V| > 1 .
	\end{cases}
	\label{equil_point}
\end{equation}
Nevertheless, even in the absence of the terms $F_1(\theta,t)$ and $Q_1(\theta,t)$, the system may or may not advance towards the stationary point.
If there is a vicinity of $(\theta_{0}, \phi_0)$ in which the system advances towards the stationary point, we say that $(\theta_{0}, \phi_0)$ is a \textit{stable point} (or \textit{stable position}), whereas if there is no such vicinity, then $(\theta_0, \phi_0)$ is an \textit{unstable} stationary point.
In Appendix~\ref{app_sec_stationary} we show that for the chosen parameters the point corresponding to $\phi_0=\pi/2$ is stable, whereas the point corresponding to $\phi_0=3\pi/2$ is unstable.
Therefore, from now on we shall discuss only the stable point $(\theta_{0}, \phi_0=\pi/2)$.

If the amplitude of the external perturbation is small, the system oscillates around the stable point.
Expanding $F_0(\theta,t)$ and $Q_0(\theta,t)$ in series around $(\theta_{0}, \phi_0)$, keeping only the linear terms, using~(\ref{exp_sin_cos_Bessel2}), and denoting $\tilde{\theta} \equiv \theta - \theta_0$, $\tilde{\phi} \equiv \phi - \phi_0$, Eqs. (\ref{dot_theta}) and (\ref{dot_phi}) become
\begin{widetext}
\begin{equation} \label{dot_theta_phi_Taylor}
\begin{cases}
	\dot{\tilde{\theta}} &\approx -  C(\theta_0) \sin(\theta_0) \Bigg\{ \alpha \sin(\theta_0) \tilde{\theta} + \tilde{\phi}
	+ \alpha\epsilon A \cos(\omega_R t )+ \alpha\epsilon \sum_{m=-\infty}^{\infty} \text{sign}^{m}(m) J_{|m|}\left(\frac{A}{\omega_R}\right) \\
	& \times \sin \Big[ (V + m \omega_R) t - k \cos(\theta_0) \Big] \Bigg\} , \\
	\dot{\tilde{\phi}} &\approx C(\theta_0) \Bigg\{\sin(\theta_0) \tilde{\theta} - [\alpha + k\epsilon \Omega_F\sin^2(\theta_0)]
	\tilde{\phi} + \epsilon A \cos(\omega_R t ) + \epsilon \sum_{m-\infty}^{\infty} \text{sign}^{m}(m)  \\
	& \times J_{|m|}\left(\frac{A}{\omega_R}\right) \sin \Big[ (V + m \omega_R) t - k \cos(\theta_0) \Big] \Bigg\} .
\end{cases}
\end{equation}
\end{widetext}

We split the system of equations~(\ref{dot_theta_phi_Taylor}) into an infinite number of systems of equations by writing
\begin{equation}
	\tilde{\theta} \equiv \tilde{\theta}_{\omega_R}+ \sum_{m=-\infty}^{\infty} \tilde{\theta}_m
	\quad {\rm and} \quad
	\tilde{\phi} \equiv \tilde{\phi}_{\omega_R}+ \sum_{m=-\infty}^{\infty} \tilde{\phi}_m
	\label{exp_phi}
\end{equation}
the solutions of the system~(\ref{dot_theta_phi_Taylor}) in linear regime are given by:
\begin{equation}
\begin{cases}
	\tilde{\theta} = &- A_{\theta\Omega} \sin \left(\omega_R t +\phi _{\omega_{R}} + \delta_{\theta \omega_R}\right)  \\
	&-\sum_{m=-\infty}^{\infty} A_{\theta m} \sin\left(\omega_m  t +\phi_{\omega_R}\right), \\
	\tilde{\phi} = &- A_{\phi\Omega} \sin \left(\omega_R t +\phi _{\omega_R} + \delta_{\phi \omega_R}\right) \\
	&-\sum_{m=-\infty}^{\infty} A_{\phi m} \sin\left(\omega_m t +\phi_{\omega_R} + \delta_{\phi m}\right),
\end{cases}
	\label{exp_phi2}
\end{equation}
where $\omega_m \equiv V+m \omega_R$ and the amplitudes are
\begin{equation} \label{A_tp_Omega}
\begin{cases}
	A_{\theta\Omega}
	&= \delta \epsilon A \dfrac{\tilde{A}_{\theta}\left(\frac{\omega_R}{\Omega_F}\right)} {f\left(\frac{\omega_R}{\Omega_F}\right)} , \\
	A_{\phi\Omega}
	&= \delta \epsilon A \dfrac{\tilde{A}_{\phi}\left(\frac{\omega_R}{\Omega_F}\right) } {\sqrt{f\left(\frac{\omega_R}{\Omega_F}\right)}} ,
\end{cases}
\end{equation}
and
\begin{equation} \label{A_tp_m}
\begin{cases}
	A_{\theta m}
	&= \epsilon \, {\rm sign}^{m}(m) J_{|m|}\left(\frac{A}{\omega_R}\right) \dfrac{\tilde{A}_{\theta}\left(\frac{\omega_m}{\Omega_F}\right)} {f\left(\frac{\omega_m}{\Omega_F}\right)} , \\
	A_{\phi m}
	& = \epsilon \, {\rm sign}^{m}(m) J_{|m|}\left(\frac{A}{\omega_R}\right) \dfrac{\tilde{A}_{\phi}\left(\frac{\omega_m}{\Omega_F}\right)} {\sqrt{f\left(\frac{\omega_m}{\Omega_F}\right)}} .
\end{cases}
\end{equation}
The functions $\tilde{A}_{\theta}$, $\tilde{A}_{\phi}$, and $f$ are defined in the Appendix~\ref{app_sec_osc_elem}.

If we define $x \equiv \omega_R/\Omega_{F}$ (in~\ref{A_tp_Omega}) or $x \equiv \omega_m/\Omega_{F}$ (in~\ref{A_tp_m}), we observe that for the chosen parameters, the ratios $\tilde{A}_{\theta}\left(x\right) / f\left(x\right)$ and $\tilde{A}_{\phi}\left(x\right) / f\left(x\right)$ have sharp maxima at $x_{res} \approx 1$, that we shall call \textit{resonances}.
These resonances are due to the fact that $f\left(x\right)$ has a minimum very close to zero, which corresponds (approximately) to $x_{res}$. So, from the equation $d f(x)/d(x^2) = 0$ we get
\begin{eqnarray}
	&& x_{res} = \bigg(\Big\{ 2 (1 - \alpha\delta \epsilon k \Omega_F) \sin^2\theta_{0} - [\alpha^2 + (\delta \epsilon k \Omega_F)^2]
	\nonumber \\
	&& \times \sin^4\theta_{0} - \alpha^2\Big\}/\Big\{2  \left(1  + \alpha^{2} + \alpha \delta \epsilon k \Omega_F \sin^{2}\theta_{0}\right)^{2}\Big\}\bigg)^{1/2}
	\nonumber \\
	&& \approx 1 \quad {\rm for\quad \alpha, \delta \epsilon k \Omega_F \ll 1}.
	\label{f_der0}
\end{eqnarray}
Therefore, the resonances appear in the system at
%
\begin{equation} \label{resonances_Omega_m}
	\begin{cases}
		\Omega_{res} = x_{res} \Omega_F \approx \Omega_{F}  \\
		|\omega_{m,res}| \equiv |V+m\Omega| = x_{res} \Omega_{F} \approx \Omega_{F} ,
	\end{cases}
\end{equation}
%
respectively.
If we denote
\begin{equation} \label{def_Ms}
	\begin{cases}
		M_{\theta} \equiv \dfrac{\tilde{A}_{\theta}\left(x_{res}\right)} {f\left(x_{res}\right)} , \\
		M_{\phi} \equiv \dfrac{\tilde{A}_{\phi}\left(x_{res}\right)} {\sqrt{f\left(x_{res}\right)}} ,
	\end{cases}
\end{equation}
the expressions of the amplitudes at the resonances (\ref{A_tp_Omega}) and (\ref{A_tp_m}) simplify to
\begin{subequations} \label{ampls_res}
\begin{equation} \label{ampls_res_Omega}
	A_{\theta\Omega} = \delta \epsilon A M_{\theta} ,
	\quad A_{\phi\Omega} = \delta \epsilon A M_{\phi} ,
\end{equation}
and
\begin{equation} \label{ampls_res_m}
	\begin{cases}
		A_{\theta m} = \epsilon \, {\rm sign}^{m}(m) J_{|m|}\left(\frac{A}{\omega_R}\right) M_{\theta} , \\
		A_{\phi m} = \epsilon \, {\rm sign}^{m}(m) J_{|m|}\left(\frac{A}{\omega_R}\right) M_{\phi} ,
	\end{cases}
\end{equation}
\end{subequations}
respectively.
We notice that the amplitude at resonance increases linearly with the amplitude of the external radiation in~(\ref{ampls_res_Omega}) and varies with $A$ as a Bessel function in~(\ref{ampls_res_m}).

\section{Discussion} \label{sec_discussion}
\begin{figure}[t]
	\centering
	\includegraphics[width=0.8\linewidth]{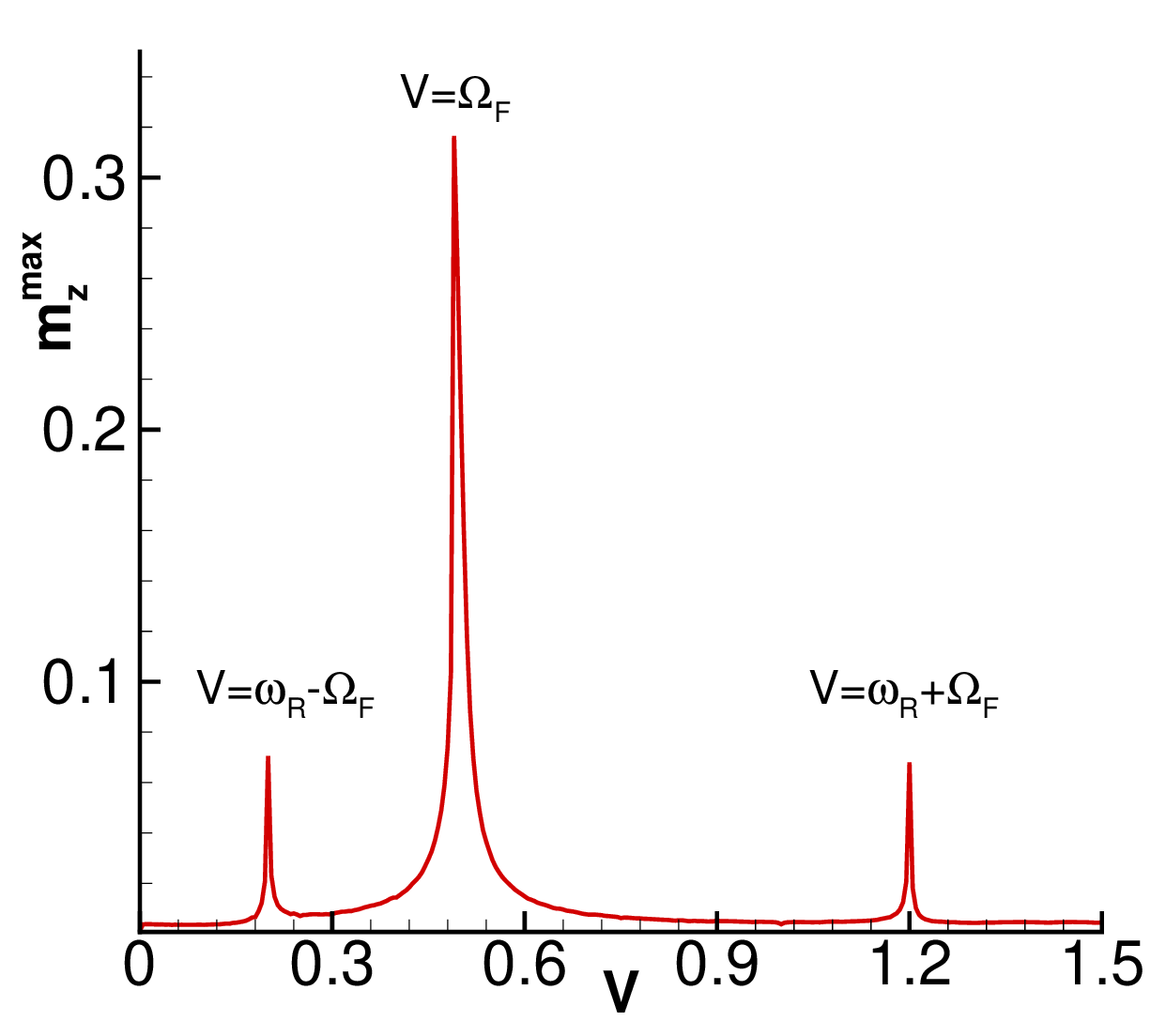}
	\caption{ Manifestation of the ferromagnetic resonance on the dependence $m_z^{max}(V)$ at $\Omega_F = 0.5, \alpha=0.001, G=0.3, k=0.01, A=0.1$ and $\omega_R=0.7$.}
	\label{fig:case1}
\end{figure}

Here we present the results of numerical simulations of the system of equations (\ref{LLG_components}) and provide a thorough comparison with the obtained analytical expressions. Figure \ref{fig:case1} shows the maximum amplitude of $m_z$ oscillations as function of $V$, for $\omega_R=0.7$.
A ferromagnetic resonance peak is observed at a voltage corresponding to the frequency of the Josephson oscillations $\omega_J = 0.5$. The results for $m_x^{max}$ and $m_y^{max}$ are qualitatively the same and are not presented.
We note that in agreement with equation (\ref{resonances_Omega_m}) for $m=\pm1$ additional peaks are observed at frequencies $\omega_R-\Omega_F$ and $\omega_R+\Omega_F$.

\begin{figure}[t]
	\centering
	\includegraphics[width=0.8\linewidth]{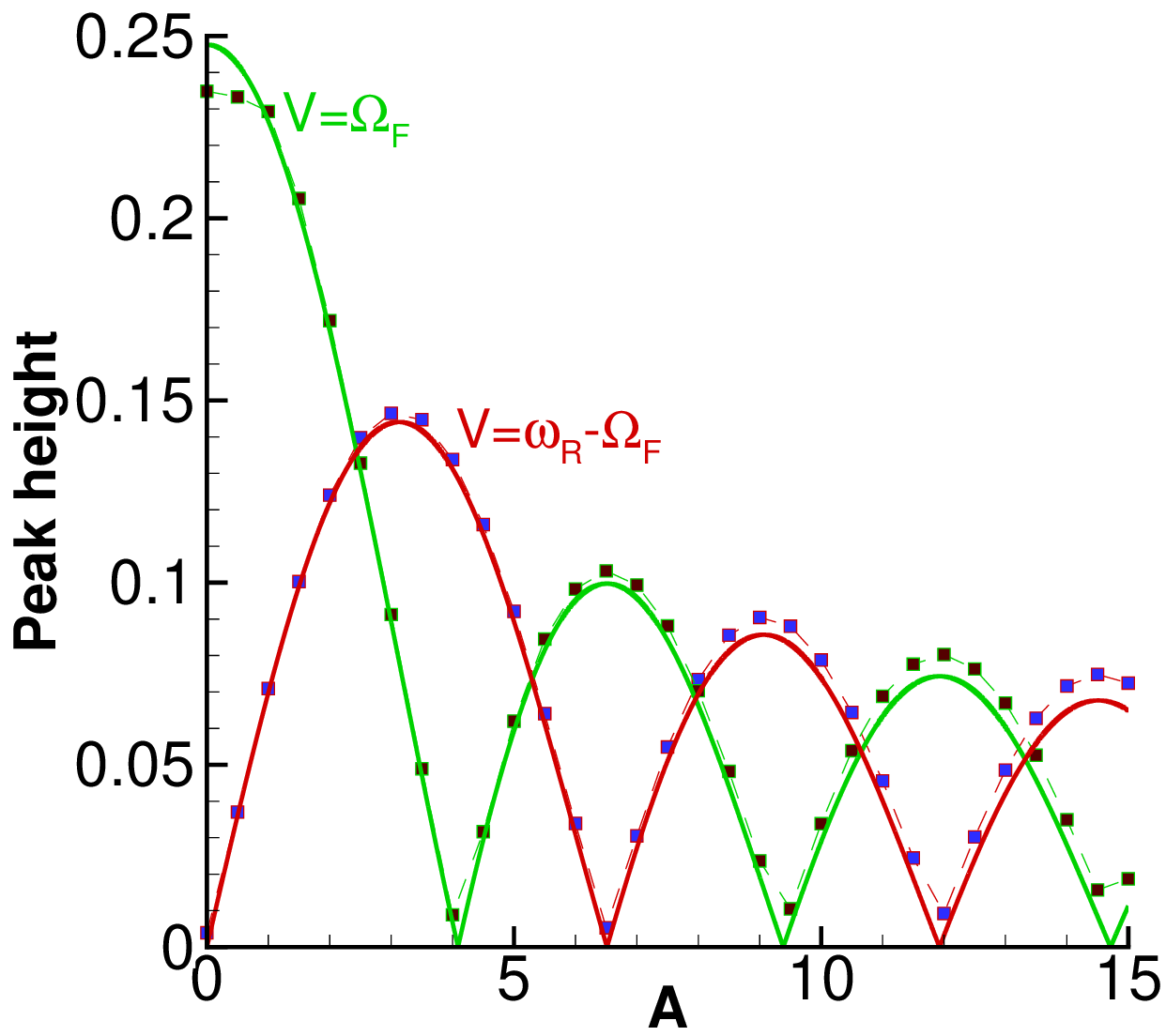}
	\caption{ (Color online) The heights of the resonance peaks as a function of amplitude $A$ at $\Omega_F = 0.5, \alpha=0.01, G=0.01, k=0.53$ and $\omega_R=1.7$. Symbols are simulated from the system of equations (\ref{LLG_components}) and solid lines are analytical approximations using (\ref{ampls_res_m}). The red lines are the height of resonance peak $V\approx\omega_R-\Omega_F$. The green lines are height of resonance peak $V\approx\Omega_F$. }
	\label{fig:3}
\end{figure}

The amplitudes of the oscillations at $V = \Omega_{F}$ and $V = \omega_R- \Omega_{F}$ are plotted in Fig.~\ref{fig:3} as functions of $A$. These choices of frequencies approximately correspond to the resonances $\omega_{0,res} = V_{res} = x_{res} \Omega_{F} \approx \Omega_{F}$ and $\omega_{-1,res} = (V-\omega_R)_{res} = x_{res} \Omega_{F} \approx \Omega_{F}$. We observe that the amplitudes approximately follow the Bessel function behaviors of~(\ref{ampls_res_m}). The symbols in Fig.~\ref{fig:3} are simulated from the system of equations (\ref{LLG_components}). Notice, that the figure demonstrates an almost perfect matching between the analytically obtained and numerically simulated results.
We note that the heights of the resonance peaks (as a functions of $A$) oscillate differently, depending on the order of the Bessel function. For example, while the height of the peak corresponding to $V \approx \Omega_{F}$ is proportional to $J_{0}\left(\frac{A}{\omega_R}\right)$, the height of the peak corresponding to $\omega_{-1,res} = (V-\omega_R)_{res} \approx \Omega_{F}$ is proportional to $J_1\left(\frac{A}{\omega_R}\right)$.
Therefore, the position of the resonances can be determined by (\ref{resonances_Omega_m}) and the their heights are determined by (\ref{ampls_res_m}).
This feature allows one to change the position and the intensity of different resonances by the amplitude and the frequency of the external periodic drive.
The first resonance condition in (\ref{resonances_Omega_m}) can be realized with variation of $\omega_R$ at fixed value of $V$.

\begin{figure}[t]
	\centering
	\includegraphics[width=0.8\linewidth]{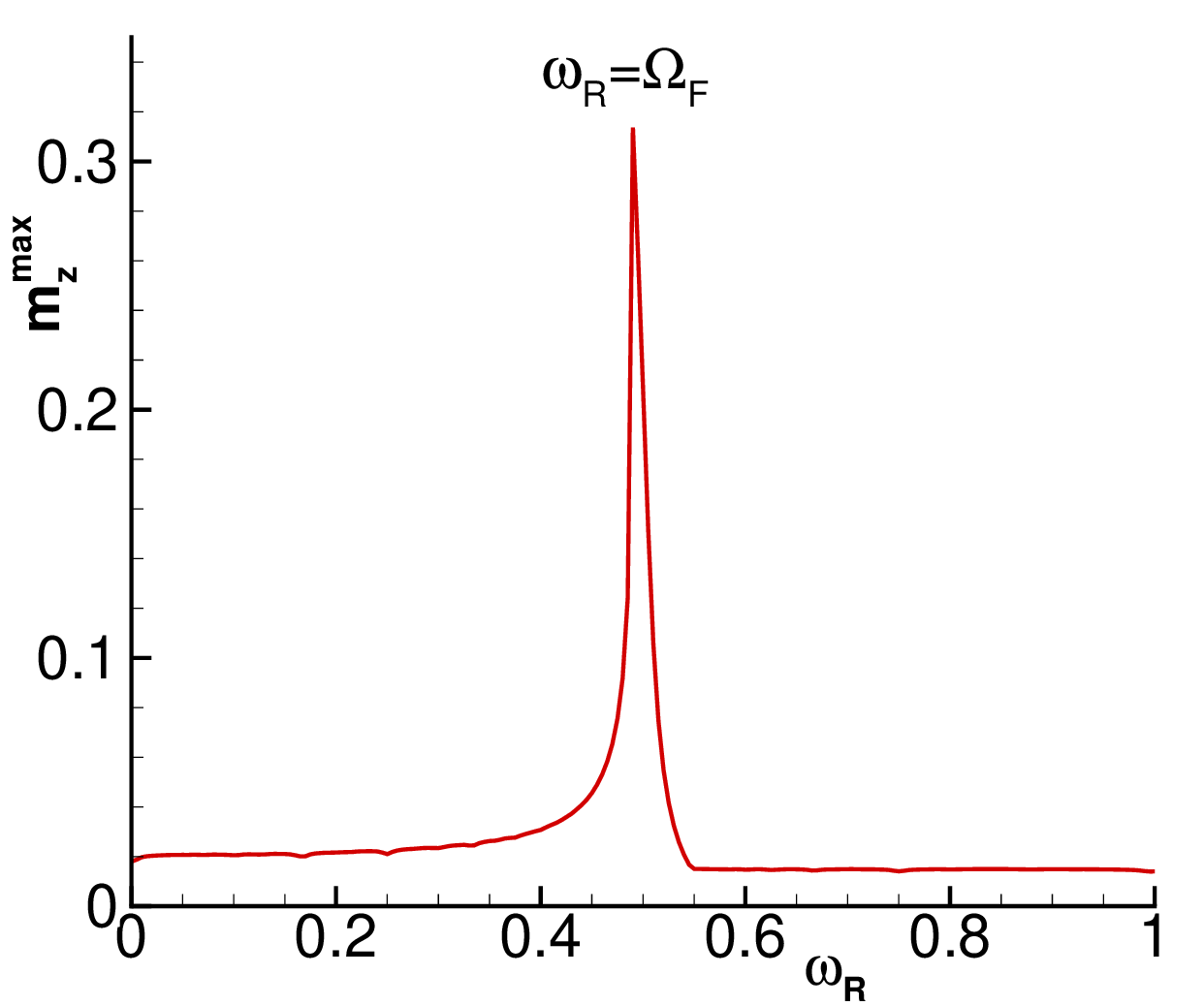}
	\includegraphics[width=0.8\linewidth]{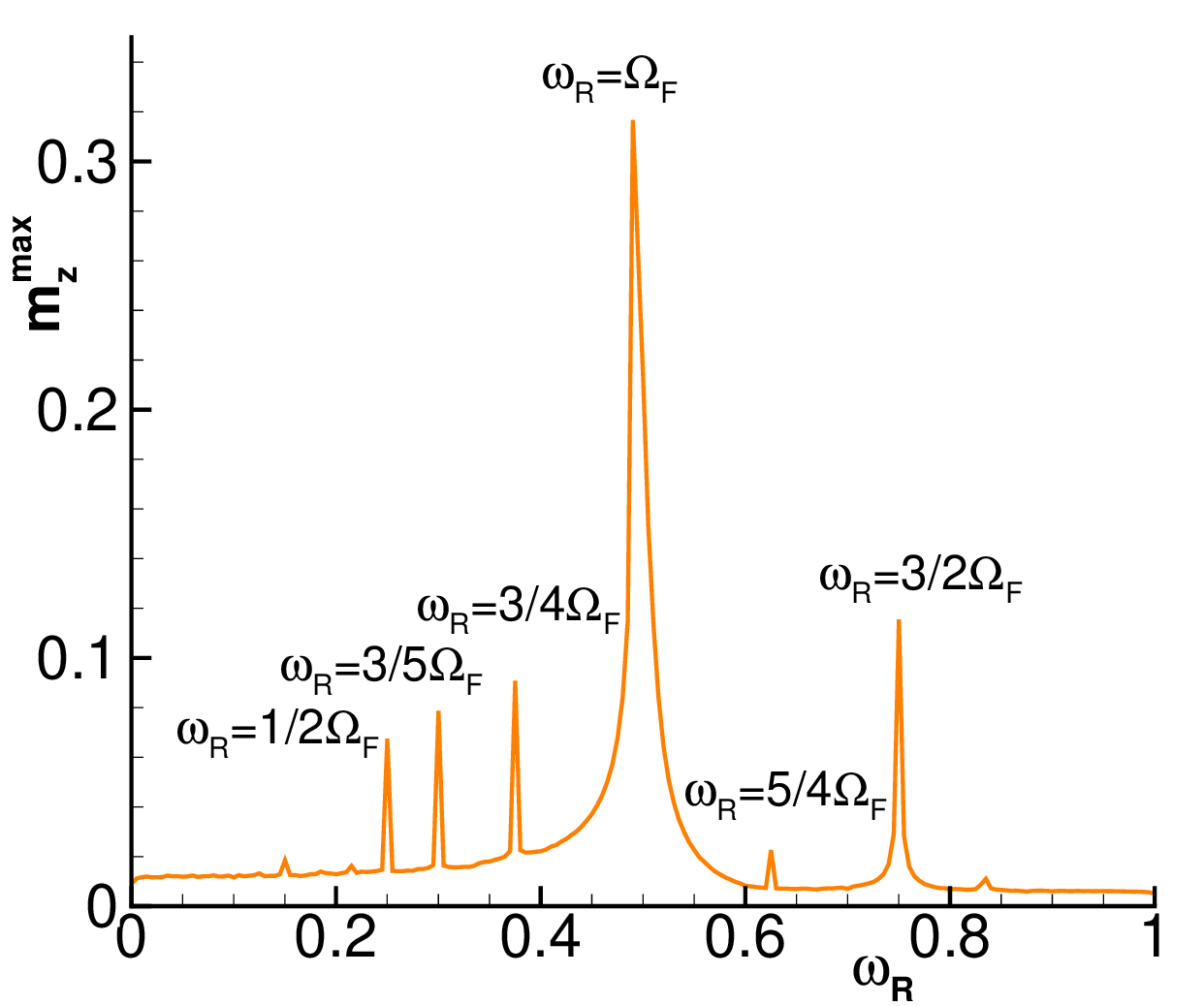}
	\caption{ Manifestation of the ferromagnetic resonance $\Omega_{F}=\omega_R\neq V$ on the dependence $m_z^{max}(\omega_R)$ at $\Omega_F = 0.5, \alpha=0.001, G=0.3, k=0.01, A=1$ and (a) $V=5$, (b) $V=2$.}
	\label{fig:4}
\end{figure}

Figure \ref{fig:4} shows the calculated maximum amplitude of oscillations $m_z^{max}$ as a function of the periodic drive frequency $\omega_R$ at different values of $V$.
A ferromagnetic resonance peak is observed (see Fig.~\ref{fig:4}(a)) at the frequency corresponding to $\omega_R\approx \Omega_{F}$ that is formed by $A_{\theta \omega_R}$~(\ref{ampls_res_Omega}).
In Fig.~\ref{fig:4}(b) we reduced the value of $V$ and additional peaks appear in the frequency interval $\omega_R \in (0,1)$, corresponding to the resonances of the amplitudes $A_{\theta m}$~(\ref{ampls_res_m}) at different values of $m$: \textbf{$(-6,-10, -5, -4, -4, -2)$} (from left to right) when the value of $V$ is reduced ($V=2$).
Note that these resonances, strongly dependent on the driving amplitudes, are similar to Kittel's ferromagnetic resonance.

The obtained results can be used to calculate the resonance condition for the $\phi_0$ Josephson junction, where the phase shift $\varphi_0$ in the current-phase relation is proportional to the magnetic moment perpendicular to the gradient of the asymmetric spin-orbit potential~\cite{buzdin08,konschelle09} \cite{smrbb-epl_18}, by setting $\delta=0$.
In this case we immediately obtain
\begin{equation}
	\left. A_{\theta \omega_R} \right|_{\delta = 0}= \left. A_{\phi \omega_R} \right|_{\delta = 0} = 0 .
	\label{A_theta_phi_Omega_d0}
\end{equation}
and
\begin{subequations} \label{tA_theta_phi_d0_max}
\begin{eqnarray}
	\left. A_{\theta m} \right|_{\delta = 0} &=&  \epsilon \, {\rm sign}^{m}(m) J_{|m|}(A/\omega_R) \frac{\sqrt{1+3\alpha^2}}{2 \alpha}
	, \label{tA_theta_d0_max} \\
	\left. A_{\phi m} \right|_{\delta = 0} &=& \epsilon \, {\rm sign}^{m}(m) J_{|m|}(A/\omega_R) \frac{\sqrt{1 - \alpha^2}}{2\alpha}
	. \label{tA_phi_d0_max}
\end{eqnarray}
\end{subequations}

Furthermore, the results presented in Ref.~\cite{snrk-jetpl_19} can be obtained at $A=0$.
In this case, only the term $m=0$ ($\omega_{m=0} = V$) survives and Eqs.~(\ref{ampls_res}) become
\begin{subequations} \label{ampls_res2}
\begin{equation} \label{ampls_res_Omega2}
	A_{\theta m}
	= \epsilon \frac{\tilde{A}_{\theta}\left(\frac{V}{\Omega_F}\right)} {f\left(\frac{V}{\Omega_F}\right)} ,
\end{equation}
and
\begin{equation} \label{ampls_res_m2}
	A_{\phi m}
	= \epsilon \frac{\tilde{A}_{\phi} \left(\frac{V}{\Omega_F}\right)} {\sqrt{f\left(\frac{V}{\Omega_F}\right) }} .
\end{equation}
\end{subequations}
The resonances can now be observed as maxima of $A_{\theta m}$ and $A_{\phi m}$ as functions of $V$ and the formulas~(\ref{resonances_Omega_m}) apply, with $V_{res} \equiv x_{res} \Omega_{F}$.

Therefore, the derived analytical description presented in Section~\ref{sec_analytic} is universal and could be applied to various systems, such as superconducting and magnetic heterostructures ($\phi_0$ Josephson junction, nanomagnet coupled to the JJ), and nonlinear pendulum.

\section{Conclusions} \label{sec_concl}

We present a thorough analytical description and classification of the possible resonances arising in the system.
The amplitudes of the induced oscillations $A_{\theta\Omega}(V, \omega_R, \Omega_{F})$ and $A_{\theta m}(\omega_m, \omega_R, \Omega_{F})$ are calculated in the linear approximation.
They have very sharp maxima (resonances) at $\Omega_{res}$ and $\omega_{m, res}$, respectively, which may be found numerically, in general.
When $\alpha, \epsilon \ll 1$, both, $\Omega_{res}$ and $\omega_{m, res}$ are approximately equal to $\Omega_{F}$.
In the limit $\alpha, \epsilon \rightarrow 0$, we have $\Omega_{res}, \omega_{m, res} \to \Omega_{F}$ and the amplitudes $A_{\theta \omega_R}, A_{\theta m}$ diverge.

We demonstrated resonance effects in a system consisting of the nanomagnet coupled to the Josephson junction, under the influence of the external periodic drive of frequency $\omega_R$.
It has been shown that the magnetic dynamics of such a system manifest additional resonances at $V+m\omega_R \approx \Omega_F$, where $m$ is an integer.
We found that the heights of the resonance peaks strongly depend on the amplitude of the external periodic drive.
Therefore, by changing the amplitude it is possible to suppress the main ferromagnetic resonance ($V=\Omega_F$) and enhance resonances at $V=|\Omega_F + m\omega_R|$.
This represents a novel method for controlling the resonance properties of the system by adjusting the driving frequency and amplitude.
In other words, by applying an external periodic drive, one can generate specific resonances, at voltages $V \approx \Omega_F +m \omega_R$, and, at the same time, suppress the FMR at $V=\Omega_F$.
This method provides a wide opportunities for experimental applications.

Our results are of practical importance.
It is known, that one of the methods for determining the characteristics of magnetic systems is FMR. The standard FMR theory based on microwave absorption in magnetic materials shows that the resonant frequency is the functions of the effective field, material and system parameters. These dependences can be used to determine the parameters of the material. On the other hand, these parameters can be varied to control the microwave absorption properties of the material. This two-way relationship between the FMR characteristics and the physical parameters of the system is usually based on analytical expressions that give the resonant frequency as a function of the material parameters (anisotropy constants, exchange and dipole couplings, etc.).
In the case of hybrid structures, however, such analytical expressions cannot be obtained and one has to resort to numerical simulation or some approximate solution.
Our results provide the necessary information for estimating the physical parameters in hybrid structures of the SFS and JJ-NM types and may have applications related to the resonance properties of hybrid structures, as well as in quantum information processing and spintronics.

\section{Acknowledgments}

The reported study was funded by the (ASRT, Egypt) - (JINR, Russia) research projects. Numerical simulations were funded by the project 22-71-10022 of the Russian Scientific Fund. Special thanks to BLTP, HybriLIT heterogeneous computing platform (LIT, JINR Russia) and Bibliotheca Alexandrina (Egypt) for the HPC servers.

\appendix

\section{The stationary point} \label{app_sec_stationary}

\begin{figure}[b]
	\centering
	\includegraphics[width=8 cm]{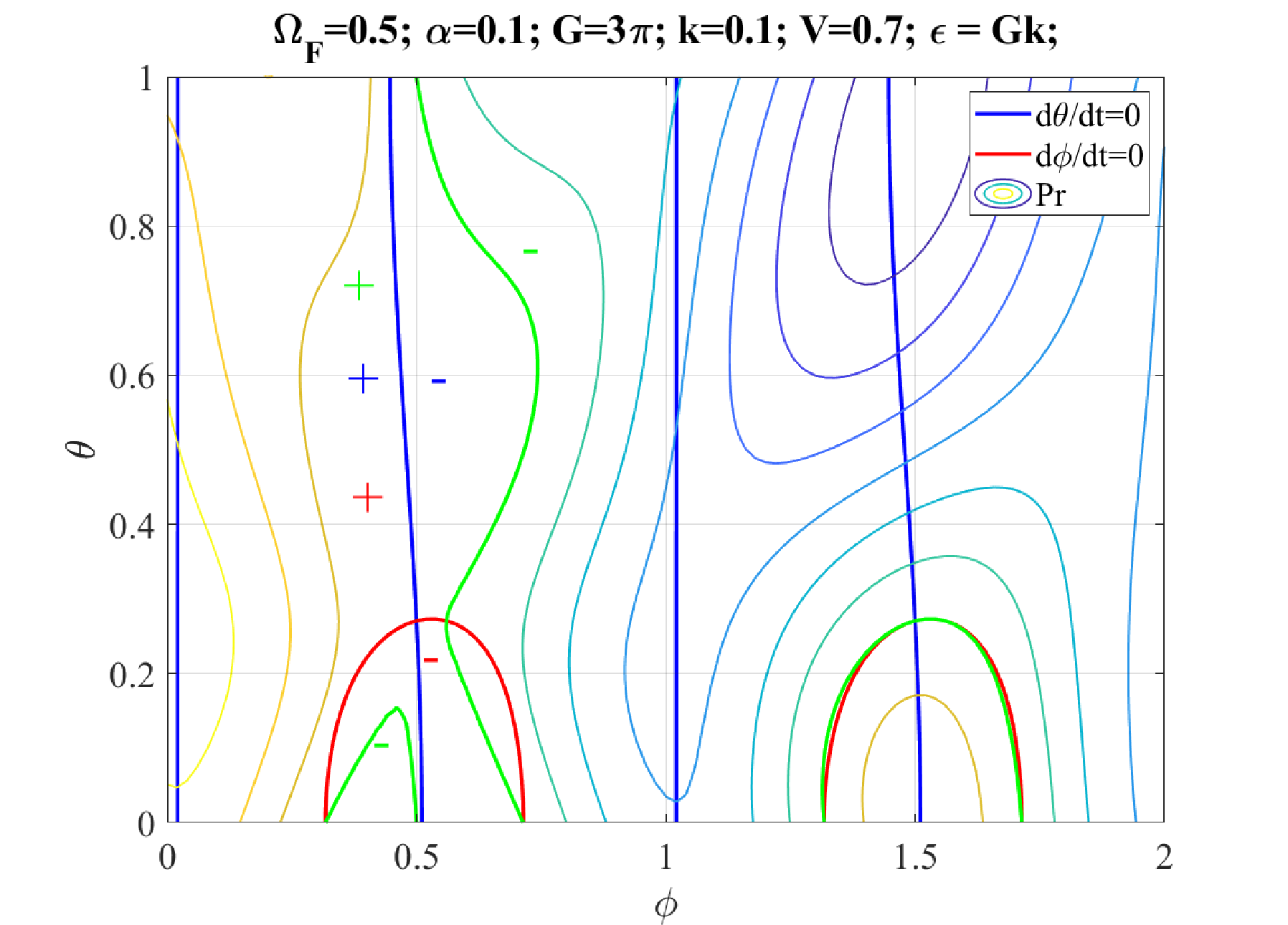}
	\caption{The contours corresponding to $F_0(\theta,\phi) = 0$ (blue thick solid line), $Q_0(\theta,\phi) = 0$ (red thick solid lines), and $P(\theta, \phi) = 0$ (green thick solid lines).
	By the colored signs $+$ and $-$ we indicate the signs of the corresponding functions: blue $+$ and $-$ signs indicate the regions in which $F_0(\theta,\phi) > 0$ and $F_0(\theta,\phi) < 0$, respectively, bordered by the thick blue solid line; red $+$ and $-$ signs indicate the regions in which $Q_0(\theta,\phi) > 0$ and $Q_0(\theta,\phi) < 0$, respectively, bordered by the thick red solid line; green $+$ and $-$ signs indicate the regions in which $P(\theta,\phi) > 0$ and $P(\theta,\phi) < 0$, respectively, bordered by the thick green solid line.}
	\label{fig_dottheta3d}
\end{figure}

To determine the stability of the stationary points~(\ref{equil_point}), we calculate the projection of the velocity $\textbf{v}\equiv(\dot{\theta}, \dot{\phi})$ on the unit vector $\hat{\textbf{s}} \equiv [(\theta_0-\theta)/u, (\phi_0-\phi)/u]$, which gives the direction from the current point $(\theta, \phi)$ to the stationary point $(\theta_{0}, \phi_0)$ (where $u = \sqrt{(\theta_0-\theta)^2 + (\phi_0-\phi)^2}$):
%
\begin{eqnarray}
	&& P(\theta, \phi) \equiv \textbf{v} \cdot \hat{\textbf{s}} = \frac{1}{\sqrt{(\theta_0 - \theta)^2 + (\phi_0 - \phi)^2}} \bigg\{ (\theta_0 - \theta)
	\nonumber \\
	&& \times \sin\theta \Big[ \alpha\epsilon\delta V - \sin\phi ( \cos\phi + \alpha \cos\theta \sin\phi) \Big] + (\phi_0 - \phi)
	\nonumber \\
	&& \times \Big[ \epsilon\delta V
	+ \Big( \delta \epsilon k \Omega_F \sin^2 \theta \cos\phi
	- \sin\phi \cos\theta + \alpha \cos\phi \Big)
	\nonumber \\
	&& \times \sin\phi \Big] \bigg\} .
	\label{eqn_convergence}
\end{eqnarray}
%
In the absence of an external time dependent perturbations (i.e., $F_1(\theta,t)=0$ and $Q_1(\theta,t)=0$), if $P(\theta, \phi) > 0$ the magnet moves towards the stationary point, whereas if $P(\theta, \phi) < 0$ the magnet moves away from the stationary point.
In Fig.~\ref{fig_dottheta3d} we see the two stationary points (the intersections of the red and blue solid lines) and observe that for the chosen parameters, the point corresponding to $(\phi_0=\pi/2)$ is in the region $P(\theta, \phi) > 0$, whereas the point $(\phi_0=3\pi/2)$ is in the region $P(\theta, \phi) < 0$.

\section{The elementary oscillations} \label{app_sec_osc_elem}

In the system~(\ref{dot_theta_phi_Taylor}) we redefine $\tilde{\theta}$ and $\tilde{\phi}$ according to~(\ref{exp_phi}).
The new variables describe ``individual'' oscillations of specified frequencies: the variables $(\tilde{\theta}_{\omega_R}, \tilde{\phi}_{\omega_R})$ for the frequency $\omega_R$ and the variables $(\tilde{\theta}_m, \tilde{\phi}_m)$ for the frequencies $\omega_m \equiv V+m\omega_R$.
These oscillations satisfy the systems of equations
\begin{equation} \label{eqs_exp_theta_phi_Omega}
	\begin{cases}
		\dot{\tilde{\theta}}_{\omega_R} & = - C(\theta_0) \sin(\theta_0) \Big\{ \alpha \sin(\theta_0) \tilde{\theta}_{\omega_R} + \tilde{\phi}_{\omega_R} \\
		& + \alpha\epsilon A \cos(\omega_R t ) \Big\} , \\
		\dot{\tilde{\phi}}_{\omega_R} & = C(\theta_0) \Big\{\sin(\theta_0) \tilde{\theta}_{\omega_R} - [\alpha + k\epsilon \Omega_F\sin^2(\theta_0)] \tilde{\phi}_{\omega_R} \\
		& + \epsilon A \cos(\omega_R t ) \Big\} ,
	\end{cases}
\end{equation}
and
\begin{equation} \label{eqs_exp_theta_phi_m}
	\begin{cases}
		\dot{\tilde{\theta}}_m & = -C(\theta_0) \sin(\theta_0) \Big\{ \alpha \sin(\theta_0) \tilde{\theta}_m + \tilde{\phi}_m
		+ \alpha \epsilon \text{sign}^{m}(m) \\
		& \times J_{|m|}\left(\frac{A}{\omega_R}\right) \sin \Big[ \omega_m t - k \cos(\theta_0) \Big] \Big\} , \\
		\dot{\tilde{\phi}}_m &= C(\theta_0) \Big\{\sin(\theta_0) \tilde{\theta}_m - [\alpha + k\epsilon \Omega_F\sin^2(\theta_0)] \tilde{\phi}_m \\
		& + \epsilon \text{sign}^{m}(m) J_{|m|}\left(\frac{A}{\omega_R}\right) \sin \Big[ \omega_m t - k \cos(\theta_0) \Big] \Big\} ,
	\end{cases}
\end{equation}
where $m \in Z$ (integer).
By this decomposition we show that the movement of the nanomagnet~(\ref{dot_theta_phi_Taylor}) is a superposition an infinite number of oscillations of frequencies $\omega_R$, and $\omega_m \equiv V+m\omega_R$.
The system (\ref{eqs_exp_theta_phi_Omega}) has solutions of the form
\begin{equation} \label{tilde_theta_phi_Omega}
	\begin{cases}
		\tilde{\theta}_{\omega_R} &
		= \frac{C_2 \left(\eta +\sqrt{\chi}+2 P \right)}{2 \sin\theta_{0}} e^{\frac{\left(\eta +\sqrt{\chi}\right) C(\theta) t}{2}}
		+ \frac{C_1\left(\eta -\sqrt{\chi}+2 P \right)}{2 \sin\theta_{0}} \\
		& \times e^{\frac{\left(\eta -\sqrt{\chi}\right) C(\theta) t}{2}} - A_{\theta\omega_R} \sin \left(\omega_R  t +\phi _{\omega_R} + \delta_{\theta \omega_R}\right) , \\ \\
		\tilde{\phi}_{\omega_R} &
		= C_2 e^{\frac{\left(\eta +\sqrt{\chi}\right) C(\theta) t}{2}}
		+ C_1 e^{\frac{\left(\eta -\sqrt{\chi}\right) C(\theta) t}{2}} \\
		& - A_{\phi\omega_R} \sin \left(\omega_R  t +\phi _{\omega_R} + \delta_{\phi\omega_R}\right) ,
	\end{cases}
\end{equation}
whereas the solutions of the system~(\ref{eqs_exp_theta_phi_m}) are
\begin{equation} \label{tilde_theta_phi_m}
	\begin{cases}
		\tilde{\theta}_m &
		= \frac{C_2 \left(\eta +\sqrt{\chi}+2 P \right)}{2 \sin\theta_{0}} e^{\frac{\left(\eta +\sqrt{\chi}\right) C(\theta) t}{2}}
		+ \frac{C_1 \left(\eta -\sqrt{\chi}+2 P \right)}{2 \sin\theta_{0}} \\
		& \times e^{\frac{\left(\eta -\sqrt{\chi}\right) C(\theta) t}{2}} - A_{\theta m} \sin\left(\omega_m  t +\phi_{\omega_R}\right) , \\
		\tilde{\phi}_m &
		= C_2 e^{\frac{\left(\eta +\sqrt{\chi}\right) C(\theta) t}{2}}
		+ C_1 e^{\frac{\left(\eta -\sqrt{\chi}\right) C(\theta) t}{2}} \\
		& - A_{\phi m} \sin\left(\omega_m t +\phi_{\omega_R} + \delta_{\phi m}\right) ,
	\end{cases}
\end{equation}
where $P \equiv \alpha + \delta k \epsilon \Omega_F \sin^2\theta_0$,
$\chi \equiv (\alpha \sin^2\theta_0 - P)^2 - 4 \sin^2\theta_0$,
$\eta \equiv - (\alpha \sin^2\theta_0 + P)$, and
$\xi_{0} \equiv C(\theta_0)^2 (P \alpha + 1) \sin^2\theta_0$ (by $\phi_{\omega_R}$ we denote an arbitrary phase).

Since $\left(\eta \pm\sqrt{\chi}\right) < 0$, the terms proportional to $C_1$ and $C_2$ decay in time and only the oscillatory terms remain.
Therefore, in the linear regime the solutions (\ref{tilde_theta_phi_Omega}) and (\ref{tilde_theta_phi_m}) could be combined and substituted into~(\ref{exp_phi}) to obtain Eqs.~(\ref{exp_phi2}), with the amplitudes~(\ref{A_tp_Omega}) and (\ref{A_tp_m}).

In Eqs.~(\ref{A_tp_Omega}) and (\ref{A_tp_m}) we have the notations
\begin{widetext}

\begin{subequations} \label{Ampl_theta_phi_Omega}
	\begin{eqnarray}
		f(x)
		&=&  \frac{ \Gamma_{1}^{2} x^{4}
			+ \left[(\alpha^2 + \Gamma_{2}^2) \sin^4\theta_0 + 2 (\alpha \Gamma_{2} - 1) \sin^2\theta_0 + \Gamma_{2}^2\right] x^2 + \sin^{4}\theta_0 }{ \Gamma_{1}^2}
		\label{def_fct_f2} \\
		\tilde{A}_{\theta} \left(x\right)
		&=& \frac{\sin\theta_0 }{\Gamma_{1}}
		\left\{\frac{\left[ \sin^2\theta_0 +  \left(\alpha^{2} \sin^2\theta_0 - 1\right) x^2 \right]^2}{\Gamma_{1}^2}
		+ \left(\frac{ \alpha + \Gamma_{2} \sin^2\theta_0}{\Gamma_{1}} + \alpha x^2 \right)^2 x^2\right\}^{1/2}
		\label{tildeA_theta_Omega_x} \\
		\tilde{A}_{\phi} (x)
		&=& \frac{1}{\left(1 + \alpha^2 + \Gamma_{2} \sin^2\theta_0\right)} x
		, \label{tA_phi_m}
	\end{eqnarray}
\end{subequations}
where $\Gamma_{1}=1 + \alpha^2 + \alpha\delta \epsilon k \Omega_F \sin^2\theta_0$, $\Gamma_{2}=\delta \epsilon k \Omega_F$. \\
In the notations~(\ref{Ampl_theta_phi_Omega}), we consider $f$, $\tilde{A}_{\theta}$, and $\tilde{A}_{\phi}$ as functions of $x$.
The phases of Eqs.~(\ref{tilde_theta_phi_Omega}) and (\ref{tilde_theta_phi_m}) are defined by the relations
\begin{eqnarray*}
	\sin\delta_{\theta\omega_R} &=& \frac{\xi_{0} \left[\xi_{0} +\frac{\omega_R^{2} \left(\alpha^{2} \sin^{2}\theta_{0}-1\right)}{\alpha  P +1}\right]}{\sqrt{\xi_{0}^2 \left[\xi_{0} +\frac{\omega_R^{2} \left(\alpha^{2} \sin^{2}\theta_{0}-1\right)}{\alpha  P +1}\right]^2
	+ C(\theta_0)^2 \omega_R^2 \left(P \xi_{0} +\omega_R^{2} \alpha \sin^{2}\theta_{0} \right)^2}} ,
	\\
	\cos \delta_{\theta\omega_R} &=& \frac{C(\theta_0) \omega_R \left(P \xi_{0} +\omega_R^{2} \alpha \sin^{2}\theta_{0} \right)}{\sqrt{\xi_{0}^2 \left[\xi_{0} +\frac{\omega_R^{2} \left(\alpha^{2} \sin^{2}\theta_{0}-1\right)}{\alpha  P +1}\right]^2
	+ C(\theta_0)^2 \omega_R^2 \left(P \xi_{0} +\omega_R^{2} \alpha \sin^{2}\theta_{0} \right)^2}} ,
	\\
	\sin\delta_{\phi\omega_R} &=& \frac{\omega_R C(\theta_0)\eta}{\sqrt{\left(\omega_R^{2}-\xi_{0} \right)^{2} + C(\theta_0)^{2} \eta^{2} \omega_R^{2}}} ,
	\quad \cos \delta_{\phi\omega_R} = - \frac{\left(\omega_R^{2}-\xi_{0}\right)}{\sqrt{\left(\omega_R^{2}-\xi_{0} \right)^{2} + C(\theta_0)^{2} \eta^{2} \omega_R^{2}}} ,
	\\
	\sin\delta_{\theta m} &=& \frac{-C(\theta_0) \omega_R \left(P \xi_{0} +\omega_R^{2} \alpha \sin^{2}\theta_{0}\right)}{\sqrt{\xi_{0}^2 \left[\xi_{0} +\frac{\omega_R^{2} \left(\alpha^{2} \sin^{2}\theta_{0}-1\right)}{P \alpha +1}\right]^2 + C(\theta_0)^2 \omega_R^2 \left(P \xi_{0} +\omega_R^{2} \alpha \sin^{2}\theta_{0}\right)^2}} ,
	\\
	\cos\delta_{\theta m} &=& \frac{\xi_{0} \left[\xi_{0} +\frac{\omega_R^{2} \left(\alpha^{2} \sin^{2}\theta_{0}-1\right)}{P \alpha +1}\right]}{\sqrt{\xi_{0}^2 \left[\xi_{0} +\frac{\omega_R^{2} \left(\alpha^{2} \sin^{2}\theta_{0}-1\right)}{P \alpha +1}\right]^2 + C(\theta_0)^2 \omega_R^2 \left(P \xi_{0} +\omega_R^{2} \alpha \sin^{2}\theta_{0}\right)^2}} ,
	\\
	\sin\delta_{\phi m} &=& \frac{\omega_R^{2}-\xi_{0}}{\sqrt{\left(\omega_R^{2}-\xi_{0} \right)^{2}+C(\theta_0)^{2} \eta^{2} \omega_R^{2}}} ,
	\quad
	\cos\delta_{\phi m} = \frac{\omega_R C(\theta_0) \eta}{\sqrt{\left(\omega_R^{2}-\xi_{0} \right)^{2}+C(\theta_0)^{2} \eta^{2} \omega_R^{2}}} .
	\end{eqnarray*}
\end{widetext}

\end{document}